\documentclass[twocolumn,superscriptaddress,showpacs,amsmath,amssymb]{revtex4}

\usepackage{graphicx}
\usepackage{dcolumn}
\usepackage{color}

\begin{document}

\title{Intruder level and deformation in {the} $SD${-}pair shell model}
\author{Yan An Luo}
\email{luoya@nankai.edu.cn}
\affiliation{Department of Physics and Astronomy, Louisiana State
University, Baton Rouge, LA 70803, USA}
\affiliation{Department of Physics, Nankai University, Tianjin, 300071,
P.R. China}
\author{Chairul Bahri}
\affiliation{Department of Physics and Astronomy, Louisiana State
University, Baton Rouge, LA 70803, USA}
\affiliation{Departemen Fisika, FMIPA, Universitas Indonesia, Depok
16424, Indonesia}
\author{Feng Pan}
\affiliation{Department of Physics and Astronomy, Louisiana State
University, Baton Rouge, LA 70803, USA}
\affiliation{Department of Physics, Liaoning Normal University,
     Dalian 116029, P. R. China}
\author{Jerry P. Draayer}
\affiliation{Department of Physics and Astronomy, Louisiana State
University, Baton Rouge, LA 70803, USA}
\date{\today}


\begin{abstract}
The influence of the intruder level on nuclear deformation is studied within
the framework of the nucleon-pair shell model truncated to an $SD$-pair
subspace.  The results suggest that the intruder level has a tendency to
soften the deformation and plays an important role in determining
the onset of rotational behavior.
\end{abstract}

\maketitle
Shell model calculations for all but light nuclei can only be carried out in
highly truncated model spaces.  Usually these are limited to a single major
valence shell, but for heavy nuclei this is often out of reach as well even when
the best of computational facilities are available.  As a result, early work --
for example, calculations based on the fermion dynamical symmetry model
\cite{fdsm} and the pseudo-SU(3) model \cite{su3} -- introduced additional
constraints; in particular, the nucleons in unique-parity levels were typically
constrained to a seniority zero configuration and as a result the dynamics was
driven by nucleons in the normal-parity levels. By-and-large, theories that have
used this assumption have not tested its validity.

In other venues, the importance of the intruder levels has received
considerable attention: their significance for correctly reproducing
available data, their role in determining the deformation of nuclei, and how
to properly incorporate them into other models have been repeatedly debated.
For example, studies in the single-shell {asymptotic} model and universal
Woods-Saxon model imply that the valence nucleons in the intruder level
contribute significantly to measurable quantities like B(E2) values
\cite{18}. Some mean-field theories claim that the particles in the intruder
level play a dominant role in determining the deformation \cite{19}. In Ref.
\cite{escher} the authors claim that both the normal and unique parity states
contribute significantly to the overall collectivity of a nuclear system.

In a nucleon-pair shell model theory truncated to a $SD$ subspace --
the $SD$-pair shell model (SDPM) \cite{npsm,sdpair} -- studies of the
so-called O(6)-limit nuclei $^{134,132}$Ba show that the effect of the
intruder level is sensitive to the structure of the single-particle
levels.  Specifically, with non-degenerate single-particle levels
and the occupancy of the intruder levels restricted to $S$ pairs only, the
collectivity of low-lying states, which is different from the O(6) symmetry
limit, does not agree with the experimental results.  {On the other hand, if the
single-particle levels are degenerate, (even though the occupancy
of the intruder level is limited to $S$-pairs only) like those of the
fermion dynamical symmetry model \cite{fdsm}, an O(6) behavior (soft
deformation) can be realized \cite{zhao,luo}.

The interacting boson model (IBM) reproduces rotational spectra in its
SU(3) limit \cite{IBM}. Since its building blocks  are $s$ and $d$ bosons that
are mapped from $S$ and $D$ nucleon pairs \cite{otsuka}, it is expected that the
SDPM should also be able to reproduce rotational spectra. It is the aim of this
paper to study the influence of the intruder (or unique parity) level -- which
has the opposite parity to the other (or normal parity) levels -- on
deformation, and thus determine how rotational spectra can be realized within
the framework of the SDPM.

The SDPM Hamiltonian is taken to be
\begin{eqnarray}
H & = & H_0 -{\frac 1 2}\kappa (Q^2_{\pi}+ Q^2_{\nu})\cdot
(Q^2_{\pi}+ Q^2_{\nu}), \label{Hamiltonian} \\
H_0 &=&\sum_{a\sigma} \epsilon_{a\sigma}n_{a\sigma},\\
Q_\sigma ^2 & = & \sum_i r^{2}_{{\sigma} i} Y^{2}(\theta_{{\sigma }i},
\phi_{{\sigma} i});
~~\sigma=\pi,~\nu,
\end{eqnarray}
where $Q^2_{\pi}$ and $Q^2_{\nu}$ are, respectively, the proton and neutron
quadrupole operators. The $E2$ transition operator is
\begin{eqnarray}
T(E2) & = & e_{\pi}Q^2_{\pi} + e_{\nu} Q^2_{\nu},
\end{eqnarray}
where $e_\pi$ and $e_\nu$ are effective charges of the proton and neutron,
respectively.

The basis of this model is constructed from collective $S$ and $D$
pairs, which are
defined as
\begin{eqnarray}
S^\dagger & = & \sum_j {\hat j}{\frac {v_j}{u_j}}(C_j^\dagger \times
C_j^\dagger)^0  \nonumber \\
D^\dagger & = & {\frac 1 2}[Q^2,~S^\dagger],
\end{eqnarray}
where ${\hat j}=\sqrt{2j+1}$, and $v_j$, $u_j$ are the occupied and
unoccupied amplitudes for orbit $j$, obtained by solving the BCS equation.
(Details can be found in Ref. \cite{npsm,sdpair}.)

To explore the influence of the intruder level on the deformation, for
simplicity, we set the normal parity levels to be degenerate and vary
the single-particle energy $\epsilon$ of the intruder level from $-0.15$ MeV to
$0.15$ MeV.  The pairing strength used in solving the BCS equation was taken to
be $0.01$ MeV for both the proton and neutron sectors of the space,
and $\kappa$ was fixed at $0.1$ MeV/$r_0^4$ with $r_0=\sqrt{\hbar/m\omega}$ the
oscillator length.  We take $N_\pi=N_\nu=3$ for this study, with protons and
neutrons occupying the same major shell except as specifically noted in the
text.

The fractional occupancies of the BCS-determined $S$ pair configuration versus
the position of the intruder level is shown in Fig. \ref{dist} for the 50-82
shell. Note that for $\epsilon=-0.15$ {MeV}, the fractional occupancy of the
intruder level is close to unity, implying that proton and neutron
$S$ pairs will primarily occupy the intruder level. As the single-particle
energy of the intruder level increases, the fractional occupancy of the intruder
level decreases. The higher the position of the intruder level relative to the
normal parity levels, the lower its fractional occupancy, and accordingly,
the less the intruder level will contribute to the overall structure of the
low-lying states.  (When $\epsilon=0.0$ MeV the levels are all degenerate and in
this case the fractional occupancy of each $j$-shell is simply the ratio
$(2j+1)/\Omega$ where $\Omega = \Sigma_j (2j+1)$ is the total degeneracy of
the system.) The behavior of the fractional occupancies for the normal parity
levels is opposite to that of the intruder level since the sum of the fractional
occupancies is normalized to unity.

As can be seen from Fig. \ref{ab_be2}, for yrast states in the 50-82
shell, the higher the relative energy of the intruder level, and hence a smaller
fraction of the protons and neutrons occupying the intruder level, the stronger
the E2 transition strengths and correspondingly the larger the deformation.
Here, the effective charges were fixed at the usual values of $e_\pi = 1.5 e$
and $e_\nu = 0.75 e$. The results for the other shells are similar and therefore
not shown.

\begin{figure}
\includegraphics[width=8cm]{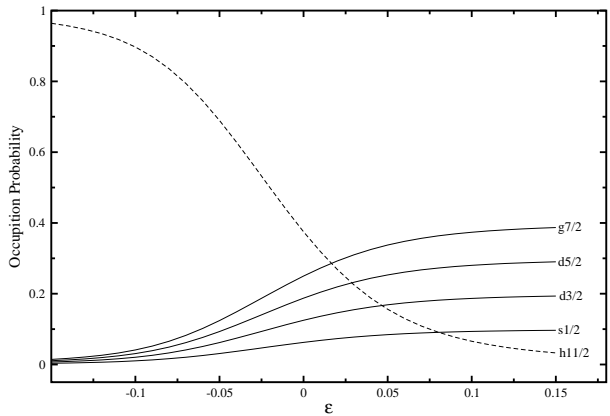}
\caption{Fractional occupancies of the 50-82 shell.
The pairing interactional strength was fixed
at 0.01 MeV in solving BCS equation.}
\label{dist}
\end{figure}

\begin{figure}
\includegraphics[width=8cm]{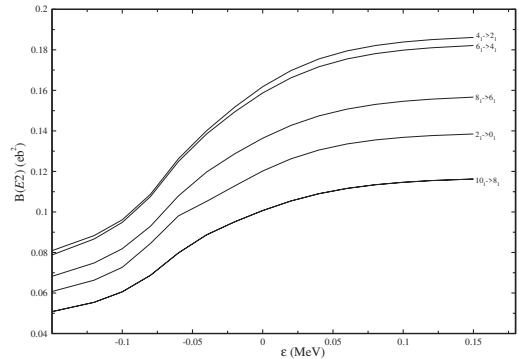}
\caption{Absolute B(E2) values for the 50-82 shell versus the
single-particle
energy of the h$_{11/2}$ intruder level.  The pairing interactional strength is
fixed at 0.01 MeV in solving BCS equation.}
\label{ab_be2}
\end{figure}

The effect of the intruder level on the rotational nature of the spectrum is
presented in Table \ref{ratio} for the calculated yrast states. In
the $ds$-shell
(the $N=Z=8$-$20$ region), the spectrum exhibits a rotational structure with
$E_{4_1^+}/E_{2_1^+}=3.3$ when the position of the intruder level is at
$\epsilon=0.07$ MeV. In general, as the intruder level moves up in energy the
rotational structure improves. For example,
$E_{4_1^+}/E_{2_1^+}=2.756$ when the
intruder and the normal parity levels are degenerate
($\epsilon_{7/2}=0$), while it
is 3.331 when $\epsilon_{f_{7/2}}=0.15$ MeV. This is counter to the
statement made in Ref. \cite{19} that the intruder levels play a dominant role
in deformation; we find that the intruder level tends to reduce deformation in
the SDPM. Specifically, the lower the intruder level, the greater the
contribution from the intruder level to the low-lying states, but the smaller
the deformation.  From Table \ref{ratio} one can
also see that the position of the intruder levels that reproduces the
rotational
$E_{4_1^+}/E_{2_1^+}=3.3$ value decreases with an increase in the number of the
single-$j$ shells and the number of single-$j$ values in each major shell. The
single-particle energies of the intruder levels that reproduce the
$E_{4_1^+}/E_{2_1^+}=3.3$ ratio is 0.6, 0.5, 0.2  and 0 MeV for the
20-50, 50-82, 82-126 and 126-184 shells, respectively.  (Note, there are no
physical systems with protons and neutrons occupying the same shell for the
82-126 and 126-184 shells.) We give the B(E2) ratios for each major shell in
Table. \ref{BE2}, from which one can see that the B(E2) ratio is close to the
SU(3) limit of an IBM theory \cite{IBM}.

\begin{table}
\caption{The energy ratio $R_J=E_{J_1^+}/E_{2_1^+}$ for each major shell.
The position of the single-particle level for the intruder level is
labeled by $\epsilon$.  The boldface entries correspond to the position where
the rotational structure, $R_2=3.3$, is realized.  {For protons and neutrons
occupying different major shells, the shells are labelled by
$\pi$ ($\nu$) for protons (neutrons) and the single-particle
energies for the intruder levels are put in a pair.}}
\label{ratio}
\begin{ruledtabular}
\begin{tabular}{ccccccc}
shell&$\epsilon$ (MeV) & J=4 & J=6 & J=8 & J=10 & J=12 \\\hline
8-20 ($ds$)
&-0.15 & 1.65 &2.86 & 3.82&  4.95&  6.06\\
& 0.00 & 2.76 &5.23 & 8.34& 12.00& 16.07\\
& \textbf{0.07} & \textbf{3.30} & \textbf{6.85}&\textbf{11.63}&
   \textbf{17.60}& \textbf{24.71}\\
& 0.15 & 3.33 &6.99 &11.97& 18.25& 25.84\\\hline
20-50
&-0.15 & 2.62& 4.00 & 5.60&  7.33&  8.20 \\
& 0.00 & 3.16& 6.37 &10.52& 15.48& 21.13\\
& \textbf{0.06} & \textbf{3.30}& \textbf{6.86} &\textbf{11.62}&
   \textbf{17.52}& \textbf{24.52}\\
& 0.15 & 3.32& 6.92& 11.78& 17.88& 25.21\\\hline
50-82
&-0.15 & 2.69& 4.77 & 7.24&  9.91& 12.56\\
& 0.00 & 3.17& 6.37 &10.51& 15.44& 21.05\\
& \textbf{0.05} & \textbf{3.30}& \textbf{6.85} &\textbf{11.59}&
   \textbf{17.45}& \textbf{24.35}\\
& 0.15 & 3.32& 6.92 &11.78& 17.86& 25.14\\\hline
82-126
&-0.15 & 2.64& 4.83 & 7.50& 10.56& 13.95\\
& 0.00 & 3.28& 6.77 &11.39& 17.03& 23.59\\
& \textbf{0.02} & \textbf{3.31}& \textbf{6.88} &\textbf{11.65}&
   \textbf{17.56}& \textbf{24.53}\\
& 0.15 & 3.33& 6.99 &11.96& 18.24& 25.82 \\\hline
126-184
&-0.15 & 2.44& 4.253 & 6.394 & 8.817& 11.436\\
& \textbf{0.00}  & \textbf{3.31}& \textbf{6.89} & \textbf{11.67}&
   \textbf{17.60}& \textbf{24.58}\\
& 0.15 & 3.33& 6.99 & 11.97&18.25&25.82\\\hline
(50-82)$_{\pi}$ & (-0.15, -0.15) & 2.65 & 4.70 & 7.31 &
10.43 & 13.94 \\
(82-126)$_{\nu}$
& \textbf{(0.04, 0.02)} & \textbf{3.30} & \textbf{6.85} & \textbf{11.62} &
\textbf{17.57} & \textbf{24.73} \\
& (0.15, 0.15) & 3.33 & 6.98 & 11.97 & 18.37 & 26.33 \\ \hline
(82-126)$_{\pi}$ & (-0.15,-0.15) & 2.47 & 4.38 & 6.71 &
9.41 & 12.36 \\
(126-184)$_{\nu}$& \textbf{(0.00, 0.00)} & \textbf{3.30} & \textbf{6.84} &
\textbf{11.56} & \textbf{17.41} & \textbf{24.38} \\
& (0.15, 0.15) & 3.34 & 7.01 & 12.04 & 18.45 & 26.31 \\ \hline

IBM & & & & & & \\
SU(3)-limit& & 3.333 & 7.0 & 12.0 & 18.333 & 26.0 \\
\end{tabular}
\end{ruledtabular}
\end{table}

\begin{table}
\caption{The relative B(E2) values with effective charges
fixed as $e_\pi=1.5e$ and $e_\nu=0.75e$. Only the results for three cases are
shown here: the first one is for the case with the lowest intruder level, the
second one is for the case that the intruder level is degenerate with
normal parity levels, while the last one is for the case with
$E_{4_1^+}/E_{2_1^+}=3.3$.}
\label{BE2}
\begin{ruledtabular}
\begin{tabular}{cccccc}
shell & $\epsilon$ & ${\frac {4_1^+\rightarrow 2_1^+}
{2_1^+\rightarrow 0_1^+}}$
&
${\frac {2_2^+\rightarrow 2_1^+} {2_1^+\rightarrow 0_1^+}}$&
${\frac {0_2^+\rightarrow 2_1^+} {2_1^+\rightarrow 0_1^+}}$\\\hline
$ds$
&-0.15 & 1.315 & 0.019  & 0.937 \\
& 0.0  & 1.350 & 0.36 & 0.053\\
& 0.06 & 1.350 & 0.004& 0.008\\\hline
20-50
& -0.15 & 1.219 & 0.004 & 0.019\\
& 0.0   & 1.347 & 0.002 & 0.005\\
& 0.06  & 1.345 & 0.04 & 0.0\\\hline
50-82
& -0.15 & 1.332 & 0.0 & 0.047\\
&  0.0  & 1.346 & 0.0 & 0.005\\
&  0.05 & 1.344 &0.004& 0.0  \\\hline
82-126
& -0.15 & 1.341 & 0.509 & 0.051\\
&  0.0  & 1.345 & 0.003 & 0.001\\
& 0.02  & 1.345 & 0.004 & 0.0\\\hline
126-184
&-0.15  & 1.286 & 1.2662& 0.0\\
&0.0    & 1.346 & 0.004 &0.001\\\hline
IBM SU(3)-limit && 1.349 & 0.0 & 0.0 \\
\end{tabular}
\end{ruledtabular}
\end{table}

This analysis suggests that the presence of the intruder level tends
to soften the deformation.  Fig. \ref{dist} shows that if the intruder level
lies low in energy relative to the normal parity levels the contribution from
the intruder level to the low-lying states is large, the expected result. It
also shows that this results in a decrease in the rotational nature of the
spectrum and hence the overall deformation.  This can be understood in two ways.
First of all, the nucleons in the intruder level cannot couple directly to those
in the normal parity levels to form positive-parity pairs because of the parity
difference, the so-called ``parity blocking effect". Simply stated, this means a
nucleon in the intruder level and one in the normal parity sector cannot bind to
one another as strongly as nucleons in the separate sectors. This coupled with
the fact that the unique parity part of the space is incomplete, lacking its
other unique parity partners so it cannot form optimal coherent pairs that favor
deformation, means that the overall deformation is reduced when pairs reside in
the unique parity sector, even though a complete unique parity shell by itself
would show stronger deformation.  In short, a consequence of the strong spin-
orbit splitting that ``isolates'' the unique parity level is that as the
contribution from the intruder level increases, the deformation decreases.

To check the validity of these observations, we can apply the logic to
well-known experimental spectra. We know, for example, that there are some $ds$
shell nuclei that show well-developed rotational bands. According to our
analysis, this suggests that the intruder level should lie high
above the active normal parity levels. Indeed, for those lower $ds$ nuclei
that show strong rotational spectra, the position of the $f_{7/2}$ level
is far removed from the most active levels, Ref.\cite{book_sp}.  But as one
moves up in mass number, the role of the $f_{7/2}$ level becomes more
important, and as the results indicate the collective rotational behavior of
upper $ds$ shell nuclei drops off.  In the upper $ds$ shell, the systems can be
viewed as holes occupying the shells but in the reverse order.  Thus, in this
picture the $f_{7/2}$ intruder level lies lowest.  For the 20-50
and 50- 82 shells there are very few nuclei that show rotational
behavior and for these nuclei the lowest high-$j$
levels is split off from the normal parity partners and the intruder levels
penetrates down among the remaining normal parity levels \cite{book_sp}.

Matters are more interesting for systems with protons filling the
50-82 shell and neutrons the 82-126 shell, or protons filling the 82-126 shell
and neutrons the 126-184 shell.  In this case there are some nuclei that show
rotational behavior, such as $^{160}$Gd.  From Table \ref{ratio} we know that
for the 50-82, 82-126 and 126-184 shells, rotational structures are predicted
even though the positions of the respective intruder levels are close to the
normal parity levels.  This is especially true for the 126-184 shell where
rotational behavior is realized when the intruder level is nearly degenerate
with the normal parity single-particle levels. This means that the predictions
of the theory are once again borne out by experiment.

In this study a simple hamiltonian was used to examine the effect of
the intruder level on deformation. From the analysis one finds that
one can give bounds on the position of the intruder level for strong rotational
spectra to occur. The position of an intruder level plays a strong role
in determining the rotational nature of the spectra and this agrees, on
average, with what is found experimentally. The SD-pair shell model
also suggests that one should not ignore intruder levels because they
have a strong influence (enhancing or reducing, depending upon the relative
position of the level) on the deformation of a system.

This work was supported in part by the U.S. National Science Foundation through
Grant Nos. 9970769 and 0140300, and partially by Education Department of
Liaoning Province (202122024).


\begin{thebibliography}{99}
\bibitem{fdsm} C. L. Wu, D. H. Feng, X. G. Chen and M. Guidry, Phys. Lett. 168B
(1986) 313; Phys. Rev. C \textbf{36} (1987) 1157.
\bibitem{su3} See, e.g, J. P. Draayer, Nucl. Phys. A \textbf{520} (1990) 259c.
\bibitem{18} K. H. Bhatt, C. W. Nestor Jr. and S. Raman, Phys. Rev. C
\textbf{46}
(1992) 164.
\bibitem{19} S. \AA berg, H. Flocard and W. Nazarewicz, Ann. Rev. Nucl. Part.
Sci. \textbf{40} (1990) 439.
\bibitem{escher} J. Escher, J. P. Draayer and A. Faessler, Nucl. Phys.
A \textbf{586} (1995) 73.
\bibitem{npsm} J. Q. Chen,  Nucl. Phys. A \textbf{626} (1997) 686.
\bibitem{sdpair} J. Q. Chen and Y. A. Luo, Nucl. Phys. A \textbf{639} (1998)
615.
\bibitem{zhao} Y. M. Zhao, N. Yoshinaga, S. Yamaji and A. Arima, Phys. Rev. C
\textbf{62} (2000) 024322.
\bibitem{luo} Y. A. Luo, J. Q. Chen, Y. C. Gao, P. Z. Ning and J. P.
            Draayer, Chin. Phys. Lett. \textbf{18} (2001) 501.
\bibitem{IBM} F. Iachello and A. Arima, \textit{The Interacting Boson
Model}, Cambridge University Press, Cambridge New York, 1987. P.174, P.60.
{\bibitem{otsuka} T. Otsuka, A. Arima and F. Iachello, Phys. Lett. \textbf{
76B}, 139; Nucl. Phys. A \textbf{309} (1978) 1.}
\bibitem{book_sp} P.
J. Brussaard and P. W. M. Glaudemans,
\textit{Shell-model
applications in nuclear spectroscopy}
(North-Holland Publishing Company,
Amsterdam,
1977).
\end{thebibliography}
\end{document}